\documentclass[aps,prl,twocolumn,amsmath,amssymb,superscriptaddress,floatfix]{revtex4}
\usepackage{graphicx}
\usepackage{graphics}
\usepackage{bm}
\usepackage[usenames]{color}
\usepackage{colordvi}
\usepackage{units}
\usepackage{bbm}
\usepackage{booktabs}
\usepackage{array}
\usepackage{placeins}
\usepackage{epsfig}
\usepackage{lscape}
\usepackage{changes}
\usepackage{braket}
\usepackage{tabularx}
\newcolumntype{L}[1]{>{\raggedright\arraybackslash}p{#1}} 
\newcolumntype{C}[1]{>{\centering\arraybackslash}p{#1}} 
\newcolumntype{R}[1]{>{\raggedleft\arraybackslash}p{#1}} 
\usepackage[colorlinks=true,linkcolor=blue,citecolor=blue,urlcolor=blue]{hyperref}

\definecolor{mymagenta}{rgb}{1.0,0.0,1.0}
\definecolor{mycyan}{rgb}{0.0,1.0,1.0}
\definecolor{myyellow}{rgb}{1.0,1.0,0.0}
\definecolor{myorange}{rgb}{1.0,0.27,0.0}

\bibstyle{apsrev.bib}

\begin{document}

\title{Quantum relaxation and metastability of lattice bosons with \\
cavity-induced long-range interactions}

\author{Benjamin Bla{\ss}}
\email{bebla@lusi.uni-sb.de}
\author{Heiko Rieger}
\email{h.rieger@physik.uni-saarland.de}
\affiliation{Theoretische Physik, Saarland University, D-66123 Saarbr{\"u}cken, Germany}
\author{Gerg\H o Ro\'osz}
\email{roosz.gergo@wigner.mta.hu}
\affiliation{Wigner Research Centre for Physics, Institute for Solid State Physics and Optics, H-1525 Budapest, P.O. Box 49, Hungary}
\affiliation{Institute of Theoretical Physics, Szeged University, H-6720 Szeged, Hungary}
\author{Ferenc Igl{\'o}i}
\email{igloi.ferenc@wigner.mta.hu}
\affiliation{Wigner Research Centre for Physics, Institute for Solid State Physics and Optics, H-1525 Budapest, P.O. Box 49, Hungary}
\affiliation{Institute of Theoretical Physics, Szeged University, H-6720 Szeged, Hungary}
\date{\today}

\begin{abstract}
The coupling of cold atoms to the radiation field within a 
high-finesse optical resonator, an optical cavity, 
induces long-range interactions which can compete with
an underlying optical lattice. The interplay between short- and long-range
interactions gives rise to new phases of matter 
including supersolidity (SS) and density waves (DW), and interesting quantum dynamics. 
Here it is shown that for hard-core bosons in one dimension the ground state phase diagram
and the quantum relaxation after sudden quenches can be calculated exactly 
in the thermodynamic limit. Remanent DW order is observed for quenches 
from a DW ground state into the superfluid (SF) phase below a dynamical transition line. After sufficiently strong SF to DW quenches beyond a static 
metastability line DW order emerges on top of remanent SF order, 
giving rise to a dynamically generated supersolid state.

\end{abstract}


\maketitle

Cold atoms offer a broad range of possibilities to investigate properties
of strongly interacting quantum many-body systems \cite{cold-atoms-review}. 
Bosonic particles in optical lattices \cite{optical-lattices} became a 
standard experimental tool to simulate the quantum mechanics of the 
Bose-Hubbard model \cite{Bose-Hubbard}, a paradigmatic theoretical 
model displaying Mott-insulating (MI) and superfluid (SF) phases\cite{Fisher1989}.
In most experiments implementing an optical lattice the atoms have
only short range interactions, but it is known that longer-range interactions lead to new phases, like supersolids and density waves, 
and interesting dynamics \cite{Lewenstein-rev,Zoller-rev}.
One way to extend the interaction range is to couple the atoms to 
an optical cavity, which can propagate interactions between atoms, making 
the interactions effectively long-ranged \citep{Science-Esslinger}.
In combination with an optical lattice \citep{Klinder2015} 
recently an experimental setup was established \cite{Landig2016}, 
in which short- and long-range interactions compete and show density 
wave (DW) and supersolid (SS) phases. Subsequent theoretical studies
using mean field theory \cite{Chen2016,Dogra2016,Niederle2016,Sundar2016,Panas2017} and Monte-Carlo simulations
\cite{Flottat2017} supported the experimental observation of these new 
phases. However, the quantum dynamics of the cold atoms in the combined
optical cavity / optical lattice is still elusive.

A widely used set-up to study experimentally and theoretically 
the non-equilibrium dynamics of a closed many body quantum system 
is a sudden quench, in which a system is prepared in its ground state 
and system parameters are rapidly set to new values \cite{quench-dynamics}. 
The central question addresses the nature of the system's stationary 
state after a long time evolution under its deterministic quantum dynamics. 
Non-integrable systems are expected to evolve into a thermalized state, 
in which local observables can be characterized by thermal expectation values
\cite{Rigol_07,Calabrese_06,Calabrese_07,Cazalilla_06,Manmana_07,
Cramer_08,Barthel_08,Kollar_08,Sotiriadis_09,Roux_09,Sotiriadis_11}
, whereas integrable systems develop into a
state described by a generalized Gibbs ensemble\cite{wouters,pozsgay,goldstein,pozsgay1,pozsgay2,essler_mussardo_manfil,ilievski1,ilievski2,doyon}.
Sudden quenches have been studied for bosons in optical lattices 
experimentally \cite{BH-dynamics-exp} and theoretically 
\cite{BH-dymamics-theory}. The underlying Hamiltonian, the
Bose-Hubbard model, is known to be non-integrable, and thus the 
dynamics expected to thermalize, but numerical studies,
comprising DMRG in one dimension \cite{DMRG}, t-VMC in 
higher dimensions \cite{tVMC} or
numerical dynamical MFT \cite{dMFT}
indicate non-thermal behaviour for strong quenches.

In this letter we analyze the effect of cavity-induced interactions on the dynamics of bosons in an optical lattice We show that for hard-core bosons 
(i.e. strong on-site repulsion allowing only single-particle occupancy)
in one dimension the ground state phase diagram
and the quantum relaxation after sudden quenches can be calculated exactly 
in the thermodynamic limit. In general the presence of long-range 
interactions renders this system non-integrable, but the cavity-induced interactions have a special form that allow for an analytical solution.
We first determine the resulting ground state phase diagram comprising
MI, DW and SF phases, and then 
study the quantum relaxation after sudden quenches starting with
DW and SF ground states, which give rise to dynamical phase transitions.
Although the ground state phase diagram does not display a SS
phase we will show that non-equilibrium states with simultaneous DW and 
SF order do exist.

Bosons in an optical lattice with cavity-induced long-range interactions
are described by an extended Bose-Hubbard model \cite{BH-Cavity1,BH-Cavity2}.
As in the experimental set-up of \cite{Landig2016} we consider the 
case in which the lattice constant of the optical lattice is half the wave length of the cavity mode. Then, for square lattices or chains, 
the Hamiltonian is given by \cite{Landig2016}
\begin{align}
\begin{split}
\hat{H}=&-\mathcal{T}\sum_{\braket{\mathbf{r},\mathbf{r}'}}\left(\hat{b}_{\mathbf{r}}^{\dagger}\hat{b}_{\mathbf{r}'}+\text{H.c.}\right)+\frac{U}{2}\sum_{\mathbf{r}}\hat{n}_{\mathbf{r}}\left(\hat{n}_{\mathbf{r}}-1\right)
\\
&-\mu\sum_{\mathbf{r}}\hat{n}_{\mathbf{r}}
-\varepsilon\frac{1}{N}\left(\sum_{\mathbf{r}\in e}\hat{n}_{\mathbf{r}}-\sum_{\mathbf{r}\in o}\hat{n}_{\mathbf{r}}\right)^2
\label{BH-Ham}
\end{split}
\end{align}
where $\hat{b}_{\mathbf{r}}^{\dagger}$ ($\hat{b}_{\mathbf{r}}$) 
are the Bose creation (annihilation) operators, 
$\hat{n}_{\mathbf{r}}=\hat{b}_{\mathbf{r}}^{\dagger}\hat{b}_{\mathbf{r}}$ the number operators, $N$ the lattice size, 
$\mathcal{T}$ the tuneling constant, $U$ the on-site repulsion, $\mu$ the chemical potential and $\varepsilon$ the strength of the infinite-range interactions induced by the cavity. The cavity-induced long-range 
interactions are represented as the square of the density wave 
order parameter $\hat{x}$
\begin{align}
\hat{x}=\frac{1}{N}\left(\sum_{\mathbf{r}\in e}\hat{n}_{\mathbf{r}}-\sum_{\mathbf{r}\in o}\hat{n}_{\mathbf{r}}\right)
\end{align}
where $e$ and $o$ stand for even and odd lattice sizes, respectively.
Within the path integral representation of the partition function 
this square appears in the exponent and can thus be linearized
by performing a Hubbard-Stratonovic transformation, introducing 
an auxiliary field. In the thermodynamic limit $N\to\infty$ this 
auxiliary field can be integrated out by a saddle point integration
\cite{remark-classical-Mukamel} yielding the effective Hamiltonian
$H(x)$ in which the term $-\varepsilon N \hat{x}^2$ 
in (\ref{BH-Ham}) is replaced by 
$-2\varepsilon Nx\hat{x}+\varepsilon Nx^{2}$
%
with $x$ the value of the auxiliary field at the saddle point given by
the self-consistency equation 
\begin{align}
x=\braket{\hat{x}}_{\text{GS}(\hat{H}(x))}\;,
\label{SK}
\end{align}
the ground state expectation value of the imbalance. The equivalence of the two Hamiltonians $\hat{H}$ and $\hat{H}(x)$ with (\ref{SK}) is valid for bipartite lattices in arbitrary dimensions in the thermodynamic limit $N\to\infty$.

We consider the large $U$-limit, excluding multiple site
occupancies, in 1d, which is experimentally realizable 
by an appropriate modification of the setup of \cite{Landig2016}.
The Hamiltonian for a system of 
length $L$ is thus given by
\begin{align}
\begin{split}
\hat{H}(x)=
&-\mathcal{T}\sum_{j=1}^L \left(\hat{a}_{j}^\dagger\hat{a}_{j+1}+
\hat{a}_{j}\hat{a}_{j+1}^{\dagger}\right)+\varepsilon L x^{2}\\
&-\sum_{j\;even}(\mu+2\varepsilon x)\hat{a}_j^\dagger\hat{a}_j 
 -\sum_{j\;odd}(\mu-2\varepsilon x)\hat{a}_j^\dagger\hat{a}_j 
\label{Ham-1d}
\end{split}
\end{align}
and can be solved analytically. $\hat{a}_{j}^{\dagger}$ ($\hat{a}_{j}$) are the hard-core Bose creation (annihilation) operators. A Jordan-Wigner transformation followed by a Fourier transformation takes the Hamiltonian to the fermionic form
\begin{align}
\hat{H}(x)=-\sum_{k>0}(c_k^\dagger,c_{k-\pi}^\dagger)
\left(
\begin{array}{cc}
\alpha_k & \gamma_k\\
\gamma_k  & \beta_k
\end{array}
\right)
\left(
\begin{array}{c}
c_k\\
c_{k-\pi}
\end{array}
\right)
+2\varepsilon x^2
\label{Eq:Hamiltonian_c_k}
\end{align}
with $k=(2n-1)\cdot\pi/L$, $n=1,2,\ldots,L/2$, 
$\alpha_k=\mu+2\mathcal{T}\cos(k)$, 
$\beta_k=\mu-2\mathcal{T}\cos(k)$ and $\gamma=2\varepsilon x$.
The $k$ and $k-\pi$ modes can be decoupled by a canonical transformation 
diagonalizing the Hamiltonian
$\hat{H}(x)=\sum_{0<k<\pi/2} 2[\Lambda_{k}\hat{\eta}_{k}^{\dagger}\hat{\eta}_{k}+\Lambda_{k-\pi}\hat{\eta}_{k-\pi}^{\dagger}\hat{\eta}_{k-\pi}+2\varepsilon x^{2}]$.
The energies of the eigenmodes are
$\Lambda_k=-\mu-\lambda_k$ and $\Lambda_{k-\pi}=-\mu+\lambda_k$
with $\lambda_k=(\mathcal{T}^{2}\cos^{2}(k)+\varepsilon^{2}x^{2})^{1/2}$,
thus $\lambda_k=\lambda_{\pi-k}$.
While $\Lambda_{k}<0$ for all $k\in(0,\pi/2)$, in an interval of $k$ the $\Lambda_{k-\pi}$-s can be positive. We characterize a given state by a wavenumber $k_m$, so that $\braket{\hat{\eta}_{k}^{\dagger}\hat{\eta}_{k}}_{{k_m}}=1$
for all $k$ and $\braket{\hat{\eta}_{k-\pi}^{\dagger}\hat{\eta}_{k-\pi}}_{{k_m}}=0$
for $k\in(0,k_{m})$ and $1$ for $k\in(k_{m},\pi/2)$. The energy per site is given by: $e(k_m)=L^{-1}\sum_{k\in\left(0,k_{m}\right)}\Lambda_{k}+
L^{-1}\sum_{k\in\left(k_{m},\frac{\pi}{2}\right)}
(\Lambda_{k}+\Lambda_{k-\pi}) +\varepsilon x^{2}$.
Using the representation
\begin{align}
\hat{x}=\frac{1}{L}\sum_{k>0}\left(\hat{c}_{k}^{\dagger}\hat{c}_{k-\pi}
+\hat{c}_{k-\pi}^{\dagger}\hat{c}_{k}\right)
\label{Eq:Imbalance_c}
\end{align}
the self-consistency equation
\begin{align}
x=\frac{\varepsilon x}{\pi}\int_{0}^{k_{m}}dk\,\frac{1}{\sqrt{\mathcal{T}^{2}\cos^{2}(k)+\varepsilon^{2}x^{2}}}
\label{Eq:Self-consistency_equation}
\end{align}
for the expectation value $x$ of the imbalance in the given state is derived. In the ground state $e_0=\min_{k_m} e(k_m)$.
 
\begin{figure}
\includegraphics{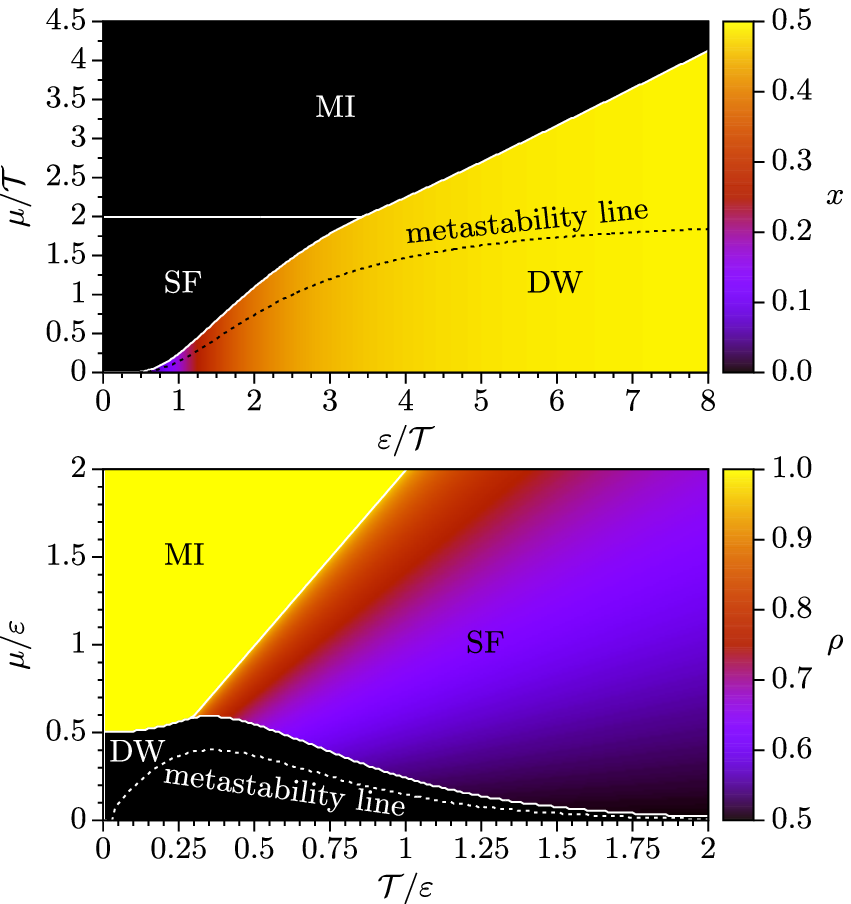}
\caption{(Colour online). Phase diagram of the one-dimensional Bose-Hubbard model with cavity-induced infinite-range interactions in the hard-core boson limit. There are three phases: Mott insulating (MI), superfluid (SF) and density wave (DW). The metastability line separates the regions with one (below) and two (above) non-trivial positive solutions of the self-consistency equation (\ref{Eq:Self-consistency_equation}) for $x$ in the DW phase.
}
\label{Fig01}
\end{figure}

The self-consistency equation always has the trivial solution $x=0$ and up to two non-trivial solutions $x\in(0,1/2]$. The stable solution minimizes the ground state energy $e_{0}$. If $x$ solves Eq.\,(\ref{Eq:Self-consistency_equation}), then $-x$ is also a solution, resulting in two equivalent ground states and thus in a broken Ising symmetry. With respect to the values of $k_{m}$ and $x$, three phases can be distinguished: Mott insulating (MI), superfluid (SF) and density wave (DW). 
\begin{table}
\renewcommand{\arraystretch}{1.5}
\begin{tabular}{ C{1.9cm} | C{1.9cm} | C{1.9cm} | C{1.9cm} }
 & MI & SF & DW \tabularnewline
\hline
$k_{m}$ & $0$ & $\in(0,\tilde{k}_{m})$ & $\pi/2$\tabularnewline
\hline
$x$ & $0$ & $0$ & $\in(0,1/2]$\tabularnewline
\hline
$\rho$ & $1$ & $\in(1/2,1)$ & $1/2$\tabularnewline
\hline
$\Delta e$ & $>0$ & $0$ & $>0$
\end{tabular}
\caption{Characterization of the Mott insulating (MI), superfluid (SF) and density wave (DW) of the ground state phases of the Hamiltonian (\ref{Ham-1d}).
Here $x$ is the imbalance (\ref{SK}), $\rho$ the particle density, $\Delta e$ the energy gap and $k_m$ the minimum wave number of the eigenmodes occupied
in the ground state.
\label{Tab01}}
\end{table}
Table \ref{Tab01} summarizes the values of $k_{m}$, the imbalance $x$, 
the density $\rho=\langle \hat{\rho} \rangle_{\text{GS}}=1/L\sum_j\langle\hat{a}_j^\dagger\hat{a}_j\rangle_{\text{GS}}$ and energy gap $\Delta e$ 
in the three phases.
The upper bound $\tilde{k}_{m}$ of $k_{m}$ in the SF phase is given by
$\tilde{k}_m=2\cdot\arctan[\exp(\pi\mathcal{T}/\varepsilon)]-\pi/2$
and $\mu/\mathcal{T}=2\cos(\tilde{k}_{m})$ 
is the meta-stability line in Fig.\,\ref{Fig01}.
As the energy gap $\Delta e$ vanishes in the SF phase, the SF ground state is a critical ground state. Due to $x\neq0$ the Ising symmetry is broken in the DW phase.
Figure \ref{Fig01} shows the phase diagram for the Hamiltonian (\ref{Ham-1d}):
The phase transition between the DW state and the MI state or the SF state is of first order. The value of $k_{m}$ and concomitantly $x$ and $\rho$ display
a discontinuity at the transition point. 
In the SF phase $k_{m}$ varies continuously and the transition between the SF and the MI states is continuous. Due to the vanishing energy gap $\Delta e$ in the SF phase, the SF correlation function $\braket{\hat{a}_{j}^{\dagger}\hat{a}_{j+r}^{\textcolor{white}{\dagger}}}_{\text{GS}}$ decays algebraically with the distance $r$, while in the DW phase the SF correlation functions decay exponentially.

Next we turn to the non-equilibrium dynamics governed by the 
Hamiltonian (\ref{Ham-1d}) and compute the time evolution of the imbalance $x(t)$ as well as the time-dependent SF correlation functions $\braket{\hat{a}_{j}^{\dagger}\hat{a}_{j+r}^{\textcolor{white}{\dagger}}}_{t}$ after a sudden quench. The system is prepared in its ground state for a given set of parameters and then driven out of equilibrium by setting
$(\mathcal{T}_{0},\mu_{0},\varepsilon_{0})\to(\mathcal{T},\mu,\varepsilon)$.

\begin{figure}
\includegraphics{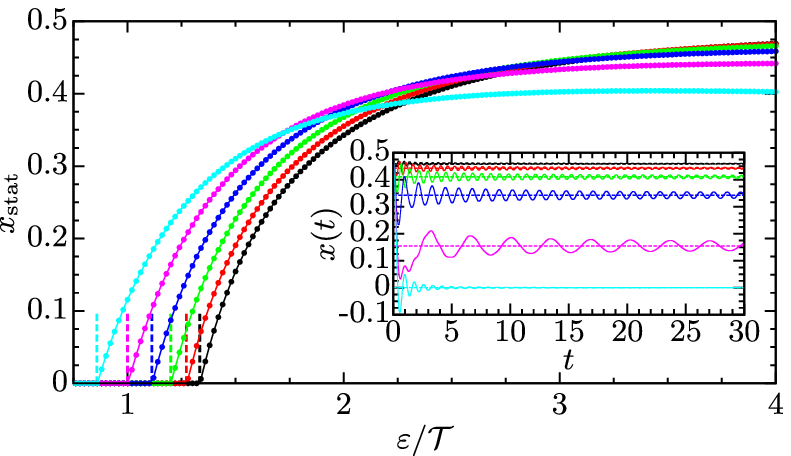}
\caption{(Colour online). \textit{Inset}: Time evolution of the imbalance $x(t)$ after quenches $\left(\mu_{0}/\mathcal{T}_{0}=1,\varepsilon_{0}/\mathcal{T}_{0}=4\right)\to\left(\mu/\mathcal{T}=1,\varepsilon/\mathcal{T}\right)$ with $\varepsilon/\mathcal{T}=3.5$ (\textcolor{black}{$\boldsymbol{-}$}), $3$ (\textcolor{red}{$\boldsymbol{-}$}), $2.5$ (\textcolor{green}{$\boldsymbol{-}$}), $2$ (\textcolor{blue}{$\boldsymbol{-}$}), $1.5$ (\textcolor{mymagenta}{$\boldsymbol{-}$}), $1$ (\textcolor{mycyan}{$\boldsymbol{-}$}). The staggered lines indicate the value $x_{\text{stat}}$ of the imbalance in the stationary state. \textit{Main panel}: Values of $x_{\text{stat}}$ after different quench protocols: $\left(\mu_{0}/\mathcal{T}_{0}=0.5,\varepsilon_{0}/\mathcal{T}_{0}\right)\to\left(\mu/\mathcal{T}=0.5,\varepsilon/\mathcal{T}\right)$ with $\varepsilon_{0}/\mathcal{T}_{0}=4$ (\textcolor{black}{$\boldsymbol{\bullet}$}), $3.5$ (\textcolor{red}{$\boldsymbol{\bullet}$}), $3$ (\textcolor{green}{$\boldsymbol{\bullet}$}), $2.5$ (\textcolor{blue}{$\boldsymbol{\bullet}$}), $2$ (\textcolor{mymagenta}{$\boldsymbol{\bullet}$}), $1.5$ (\textcolor{mycyan}{$\boldsymbol{\bullet}$}). The staggered lines indicate the dynamic phase transition according to (\ref{Eq:Dynamic_phase_transition_DW_SF}).}
\label{Fig02}
\end{figure}

The time evolution operator $\exp(-\imath\hat{H}t)$, with $\hat{H}$ given by (\ref{Ham-1d}),
can for infinitesimal time-steps $t$ be treated in the same way as the 
partition function above, resulting in an effective time-dependent 
Hamiltonian $\hat{H}(t)$ describing the dynamics. $\hat{H}(t)$ is
identical with the Hamiltonian $\hat{H}$ from Eq.\,(\ref{Ham-1d})
but with $x$ replaced by a time-dependent function $x(t)$ that
fulfils at each time $t$ the self-consistency equation 
$x(t)=\langle \psi_0 | \hat{x}(t) | \psi_0\rangle$,
where $\hat{x}(t)$ is the operator $\hat{x}$ from 
Eq.\,(\ref{Eq:Imbalance_c}) in the Heisenberg picture.
This also guarantees that the total energy is conserved under the time evolution, i.e. $\partial\braket{\hat{H}(t)}/\partial t=0$. 
The representation (\ref{Eq:Hamiltonian_c_k}) is then used to derive equations
of motion for $\hat{c}_{k}^{\dagger}(t)$ and $\hat{c}_{k-\pi}^{\dagger}(t)$ and their Hermitian adjoints in the Heisenberg picture. Expressing them
in the free fermion operators that diagonalize the momentary 
Hamiltonian $\hat{H}(t)$ yields a system of 2$L$ coupled ordinary first 
order differential equations for the time-dependent Bogoliubov-parameters.
Their time-derivative also depends on $x(t)$, 
which is determined with the time-dependent version of 
(\ref{Eq:Imbalance_c}) involving again the Bogoliubov-parameters. 
The non-linear system of ordinary differential equations is then integrated numerically using standard methods \cite{remark0}.
\begin{figure}
\includegraphics{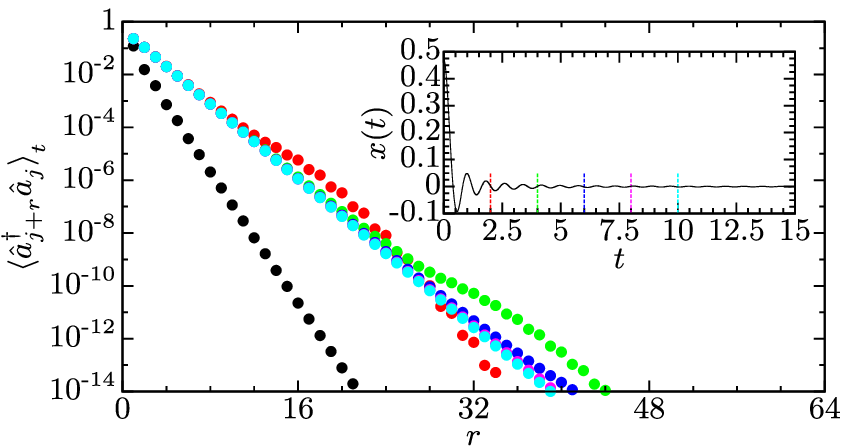}
\caption{(Colour online). Time evolution of the SF correlation function $\braket{\hat{a}_{j+r}^{\dagger}\hat{a}_{j}^{\textcolor{white}{\dagger}}}_{t}$ 
({\it main panel}) and the imbalance $x(t)$ ({\it inset})
after the quench $\left(\mu_{0}/\mathcal{T}_{0}=1,\varepsilon_{0}/\mathcal{T}_{0}=4\right)\to\left(\mu/\mathcal{T}=1,\varepsilon/\mathcal{T}=1\right)$ from the DW into the SF phase. Colour code: $t=0$ (\textcolor{black}{$\boldsymbol{\bullet}$}), $2$ (\textcolor{red}{$\boldsymbol{\bullet}$}), $4$ (\textcolor{green}{$\boldsymbol{\bullet}$}), $6$ (\textcolor{blue}{$\boldsymbol{\bullet}$}), $8$ (\textcolor{mymagenta}{$\boldsymbol{\bullet}$}), $10$ (\textcolor{mycyan}{$\boldsymbol{\bullet}$}). The exponential decay with $r$ is preserved under the dynamic phase transition. The correlation length changes during the relaxation process to a larger value in the stationary state.}
\label{Fig03}
\end{figure}

\begin{figure}
\includegraphics{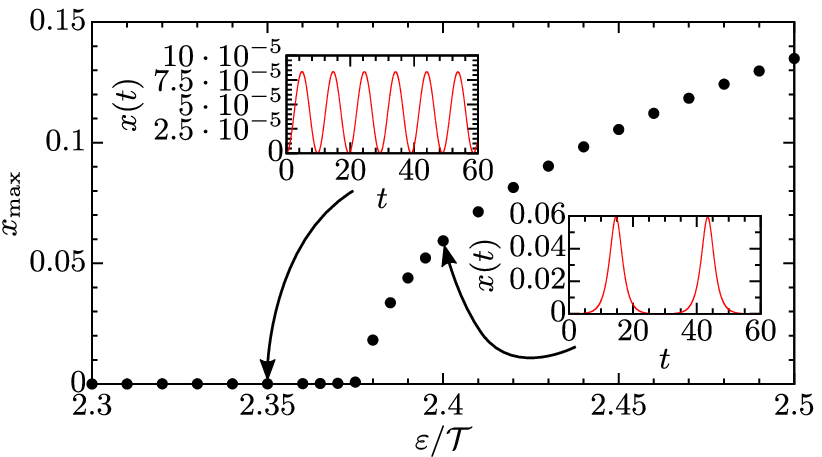}
\caption{(Colour online). \textit{Main panel}: $\varepsilon/\mathcal{T}$-dependence of $x_{\text{max}}$ in the vicinity of the dynamic phase transition for quenches with $\left(\mu_{0}/\mathcal{T}_{0}=1,\varepsilon_{0}/\mathcal{T}_{0}=0\right)\to\left(\mu/\mathcal{T}=1,\varepsilon/\mathcal{T}\right)$. The dynamic phase transition is indicated by an increase of $x_{\text{max}}$. \textit{Insets}: Time evolution of the imbalance for the quench protocols with $\varepsilon/\mathcal{T}=2.35$ (below the dynamic phase transition) and $\varepsilon/\mathcal{T}=2.4$ (above the dynamic phase transition). The value of the perturbation $x_{0}$ is $10^{-6}$.}
\label{Fig04}
\end{figure}

In the following we focus on quenches which change $\varepsilon$  and keep
$\mathcal{T}$ and $\mu$ constant \cite{remark1}.
For quenches starting in the DW phase the imbalance does not vanish in the initial state. If the strength of the infinite-range interactions is decreased $x(t)$ shows a fast decay after the quench and reaches a stationary state with only small oscillations around its stationary value $x_{\text{stat}}$ (inset of Fig.\,\ref{Fig02}). The value of $x_{\text{stat}}$ is shown in the main panel of Fig.\,\ref{Fig02} for different quench protocols. The dynamic phase transition between the DW and the SF phase occurs when $x_{\text{stat}}$ goes to $0$, which is the case at
\begin{align}
(\varepsilon/\mathcal{T})_{\text{crit}}=\frac{2\varepsilon_{0}/\mathcal{T}_{0}}{2+\varepsilon_{0}/\mathcal{T}_{0}}
\label{Eq:Dynamic_phase_transition_DW_SF}
\end{align}
independently of the chemical potential $\mu$.
We checked whether superfluidity emerges after a DW to SF quench by 
calculating the time-dependent SF correlation functions $\braket{\hat{a}_{j+r}^{\dagger}\hat{a}_{j}^{\textcolor{white}{\dagger}}}_{t}$. The result is shown 
Fig.\,\ref{Fig03}, showing that the correlations decay exponentially after
the quench implying the absence of SF order also for $x_{\text{stat}}=0$.

For quenches starting in the SF phase it is $x(t=0)=0$ which implies that
$x(t)=0$ is a solution of the self-consisteny equation for all $t>0$. 
In the following we test the stability of this solution adding a small perturbation $x_{0}$ to the imbalance. We find that for quenches not too deep into the DW phase $x(t)$ remains close to $0$ (left inset of Fig.\,\ref{Fig04}). However, for stronger quenches on the other hand the value $x_{\text{max}}$ of the maxima in the oscillations of 
$x(t)$ becomes much larger than the perturbation $x_{0}$ (right inset of Fig.\,\ref{Fig04}). We can identify a sharp dynamical transition between a region for which the solution $x(t)=0$ is stable and a region for which it is not stable (main panel of Fig.\,\ref{Fig04}). This dynamical transition coincides with the metastability line within the DW phase and thus depends on the chemical potential in contrast to the dynamical phase transition after quenches from the DW into the SF phase. Denoting the distance to the dynamic phase transition with $\delta$, i.e. $\delta=(\varepsilon/\mathcal{T})-(\varepsilon/\mathcal{T})_{\text{crit}}$, we find power-law dependences $x_{\text{max}}\propto\delta^{1/2}$ and $t_{\text{max}}\propto\delta^{-1/2}$ for small values of $\delta$ with $x_{\text{max}}$ the value of the maxima of the oscillations of $x(t)$ and $t_{\text{max}}$ the time of the first maximum of $x(t)$ after the quench.

\begin{figure}
\includegraphics{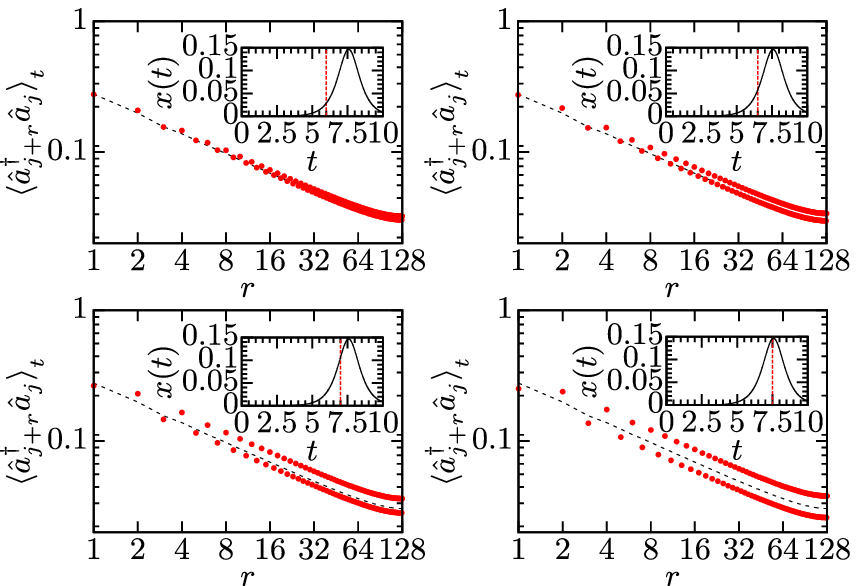}
\caption{(Colour online). Time evolution of the SF correlation function $\braket{\hat{a}_{j+r}^{\dagger}\hat{a}_{j}^{\textcolor{white}{\dagger}}}_{t}$ after the quench $\left(\mu_{0}/\mathcal{T}_{0}=1,\varepsilon_{0}/\mathcal{T}_{0}=0\right)\to\left(\mu/\mathcal{T}=1,\varepsilon/\mathcal{T}=2.5\right)$ from the DW into the SF phase. The algebraic decay of the SF correlation function in the initial SF ground state (scattered black line) is preserved under the dynamic phase transition, but for $x(t)>0$ the curve splits up into two curves for even and odd distances between the spins.}
\label{Fig05}
\end{figure}

For quenches from the SF phase into the DW phase across the metastability line we find that the SF correlation functions still decay algebraically indicating
the simultaneous presence of quasi-long-range SF order and DW order, see 
Fig.\,\ref{Fig05}. Consequently the system attains dynamically
generated supersolid (SS) properties after strong enough quenches from 
the DW into the SF phase, which is not the ground state, but 
a high energy state. Furthermore it is interesting to note that for times with $x(t)>0$ there is an even-odd modulation of the SF correlation functions which increases with $x(t)$ and which disappears when the imbalance goes back to $0$ (see Fig.\,\ref{Fig05}). The density modulations reflect even-odd modulations of the SF correlation functions.

To summarize we have solved exactly the statics and non-equilibrium dynamics 
of a system of hard-core bosons in 1d with cavity-induced 
long-range interactions. Although the ground state 
phase diagram does not display a supersolid region, i.e. a phase with
simultaneous SF and DW order, we have shown that high energy states 
can dynamically generate 
(quasi-)supersolid order with periodically modulated DW and SF correlations
in the stationary state. Our predictions can be tested experimentally
by an appropriate modification of the cavity-setup used in \cite{Landig2016}
and recording the time evolution as in \cite{Kaufmann-etal}.
Since the system turns out to be integrable in 
the thermodynamic limit and therefore does not thermalize it is an 
intriguing question whether soft-core lattice bosons in a cavity in 
one or higher dimension show similar emergent density modulations 
together with sustained superfluidity after quenches from the SF phase 
into the DW phase. In the affirmative case it would indicate the absence of 
thermalization in a non-integrable system after specific quenches from the
SF phase into the DW phase.

\begin{acknowledgments}
This work was supported by the National Research Fund under Grants No. K109577 and No. K115959. H.R. extends thanks to the "Theoretical Physics Workshop" and F.I. and G.R. to the Saarland University for supporting their visits to Budapest and Saarbr\"ucken, respectively.
\end{acknowledgments}


\begin{thebibliography}{99}
\bibitem{remark0}
We used $L=256$ and $L=512$ and checked that the results we show do 
not depend on system size for the times we considered.
\bibitem{remark1}
Since the term $-\mu L\hat{\rho}$  commutes with $H(t)$ quenches
changing only the chemical potential leave the system in an eigenstate of 
$H(t)$, not necessarily the ground state, and the time evolution is trivial.
\end{thebibliography}
\end{document}